\pdfoutput=1
\documentclass[journal,comsoc]{IEEEtran}

\usepackage{xurl}
\PassOptionsToPackage{hyphens}{url}
\usepackage[colorlinks]{hyperref}

\usepackage{listings}
\usepackage{soul}
\usepackage{pgf}
\usepackage{multirow}
\usepackage{subfigure}
\usepackage{gensymb}
\usepackage[font=footnotesize,format=plain,labelformat=simple,labelsep=period]{caption}
\usepackage{balance}
\usepackage{amsmath}
\usepackage{amssymb}

\usepackage{cite}

\definecolor{dkgreen}{rgb}{0,0.6,0}
\definecolor{gray}{rgb}{0.5,0.5,0.5}
\definecolor{mauve}{rgb}{0.58,0,0.82}

\usepackage{letltxmacro}

\makeatletter
\def\blfootnote{\xdef\@thefnmark{}\@footnotetext}
\makeatother

\def\ver{2}

\def\StartVersionOne{\ifnum\ver=1}

\def\StartVersionTwo{\ifnum\ver=2}

\DeclareUnicodeCharacter{02B9}{'}

\newcommand{\blind}[1]{\ifnum\ver=1 #1\fi}
\newcommand{\normal}[1]{\ifnum\ver=2 #1\fi}

\begin{document}


\title{LEO Satellite Network Access in the Wild:\\Potentials, Experiences, and Challenges}

\author{Sami Ma, Yi Ching Chou, Miao Zhang, Hao Fang, Haoyuan Zhao, Jiangchuan Liu,~\IEEEmembership{Fellow,~IEEE}, and William I. Atlas
\thanks{Published in IEEE Network DOI: 10.1109/MNET.2024.3391271 \url{https://ieeexplore.ieee.org/document/10505311} (Apr. 19, 2024)}
\thanks{Sami Ma, Yi Ching Chou, Miao Zhang, Hao Fang, Haoyuan Zhao and Jiangchuan Liu are with the School of Computing Science, Simon Fraser University, Burnaby, BC, Canada (E-mail: \{masamim, ycchou, mza94, fanghaof, hza127, jcliu\}@sfu.ca).}
\thanks{William I. Atlas is with the Pacific Salmon Foundation (E-mail: watlas@wildsalmoncenter.org).}
}

\maketitle
\thispagestyle{plain}
\pagestyle{plain}

\begin{abstract}
In the past three years, working with the Pacific Salmon Foundation and various First Nations groups, we have established Starlink-empowered wild salmon monitoring sites in remote Northern British Columbia, Canada. We report our experiences with the network services in these challenging environments, including deep woods and deep valleys, that lack infrastructural support with some close to Starlink's service boundary at the far north. We assess the portability and mobility of the satellite dishes and the quality of existing network access in underdeveloped countries that Starlink expects to cover. Our experiences suggest that network access based on LEO satellite constellations holds promise but faces hurdles such as energy supply constraints and environmental factors like temperature, precipitation, and solar storms. The presence of wildlife and respecting local residents' culture and heritage pose further complications. We envision several technical solutions addressing the challenges and believe that further regulations will be necessary.

\end{abstract}


\section{Introduction}

Operating at around 180 km to 2,000 km above the Earth surface, Low Earth Orbit (LEO) satellites offer shorter latency and higher throughput for space-ground communications compared to traditional Geosynchronous orbit (GEO) satellites at around 35,780 km.\footnote{\url{https://earthobservatory.nasa.gov/features/OrbitsCatalog}, [Online; accessed Sep
27, 2023]} Though each LEO satellite only covers a small portion of Earth's surface, they can collaboratively form a {\em LEO Satellite Network} (LSN) constellation to achieve global coverage with handovers across satellites. LSN has been advocated as a key infrastructure for truly ubiquitous coverage in the forthcoming 6G. We have seen commercial deployment of LSN constellations in the past decade with rapidly growing attention from the general public. One of the industrial leaders, \emph{OneWeb}, is building a constellation of 648 broadband LEO satellites, which would eventually expand to 7,000.\footnote{\url{https://oneweb.net/resources/oneweb-streamlines-constellation}, [Online; accessed Sep
27, 2023]} Its rival, \emph{SpaceX's Starlink}, has launched more than 2,000 satellites to LEO and has received approval from the Federal Communications Commission (FCC) to bring that number up to 12,000.

Preliminary user experiences of LSNs, based on various reports and our own experiments with Starlink~\cite{ma2023network}, have been generally positive in urban cities. Its potentials and challenges in the wild, however, remain unclear. In the past three years, working with the Pacific Salmon Foundation, Wild Salmon Center, and various First Nations groups, we have established Starlink-empowered wild salmon monitoring sites in remote areas of Northern British Columbia. In this article, we report our experience in these challenging environments, including deep woods and deep valleys, all of which lack infrastructural support and some are at the far north, being close to Starlink's service boundary. We have also assessed the portability and mobility of dishes, along with the quality of existing networks in underdeveloped regions targeted for coverage by Starlink (Africa, in particular).  Our findings indicate that Starlink effectively supports our monitoring systems; however, energy supply poses a significant bottleneck in wild environments. Environmental variables like temperature, precipitation, and solar storms also have noteworthy impacts. Co-existing with wildlife and respecting local culture and heritage present additional challenges.

We envision several technical solutions that address infrastructural challenges, including cross-orbit collaboration integrating multi-tier space networks, coordinated multi-path transmission with inter-satellite-links, and mobile edge computing anchored at LEO satellites. As LSNs rapidly develop, we anticipate achieving global, anytime networking in the near future. Yet, challenges related to the environment, social, and culture will persist, necessitating appropriate regulations.

\section{Starlink in the Wild: Experiences and Lessons}
\label{sec:starlink_in_wild}

Starlink's network coverage now spans North America—with intermittent service in the northernmost areas-as well as Western Europe, selected regions in Australia, and New Zealand.\footnote{\url{https://satellitemap.space}, [Online; accessed Sep 27, 2023]} To understand the Starlink's performance, we first established testing sites in urban centers of Canada and the United States using first-generation dishes. These sites served as vantage points for communicating with globally-distributed servers. We compared the results against those achieved using traditional terrestrial networks between the same source-destination pairs. 

Our findings reveal that Starlink's average end-to-end latency slightly exceeds terrestrial networks by $10\%$, with a significantly higher standard deviation of $380\%$, indicating greater latency variability. While Starlink outperforms traditional GEO satellites in throughput, it falls short of matching terrestrial network capacities and exhibits high asymmetry. Specifically, TCP download speeds range from 66 to 108 Mbps for Starlink, considerably lower than the 238 to 799 Mbps range for terrestrial networks. Conversely, Starlink's upload throughput ranges from 6 to 7 Mbps, compared to 38 to 64 Mbps for terrestrial networks\cite{ma2023network}.

From end users' view, Starlink offers a competitive, albeit not fully equivalent, alternative to terrestrial networks in urban areas with extensive satellite coverage. It complements existing terrestrial infrastructure and can potentially serve as a substitute in specific scenarios.

\subsection{Remote Salmon Monitoring: A Satellite Use Case}

\begin{figure}[t]
    \centering
    \includegraphics[width=.9\linewidth]{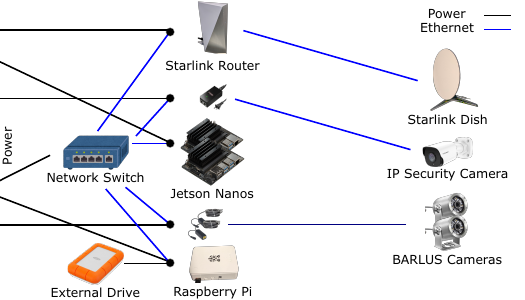}
    \caption{Salmon monitoring system setup diagram.}
    \label{fig:salmon_monitoring_system}
\end{figure}

Focusing on remote regions, satellite communication systems, such as Starlink, address the challenge of limited or costly network services in wild and isolated environments. However, some northern areas experience degraded performance at the edge of the LEO satellite constellation. In collaboration with the Pacific Salmon Foundation, Wild Salmon Center, and various First Nations groups in British Columbia, Canada, we deployed real-time salmon monitoring systems using Starlink at the KwaKwa Creek and Koeye River, as shown in Fig.~\ref{fig:salmon_monitoring_system}. Our system, utilizing deep learning for automated counting, relies on Starlink as a backbone for remotely monitoring underwater cameras and managing on-site microprocessors. Two types of microprocessors are employed: Raspberry Pis for remote access and pipeline management, and Nvidia Jetson Nanos for running the deep learning model to track salmon from each underwater camera. The Jetsons retrieve full-resolution video data via the Real-Time Streaming Protocol (RTSP), storing it on an external hard drive for manual retrieval, while the Raspberry Pi uploads a lower-quality video stream to Amazon Web Services (AWS) Kinesis Video Streams (KVS) for basic remote monitoring.\footnote{\url{https://docs.aws.amazon.com/kinesisvideostreams/latest/dg/what-is-kinesis-video.html}, [Online; accessed Sep 27, 2023]}

Starlink surpasses Xplornet, a GEO satellite-based Internet provider that was used in these remote areas, with superior service. Xplornet's download and upload maximum throughputs are capped at 25 Mbps and 1 Mbps, respectively, with a yearly data cap of 250 GB, while often delivering unreliable connectivity.\footnote{\url{https://www.xplore.ca/shop/internet-packages/satellite/}, [Online; accessed Sep 27, 2023]}

\subsubsection{Northern Shoreline}

\begin{figure}[t]
    \centering
    \subfigure[Dish at the estuary of the Koeye River towards a clear sky.]{
        \includegraphics[trim={0 0.5cm 0 0.6cm},clip,width=.45\linewidth]{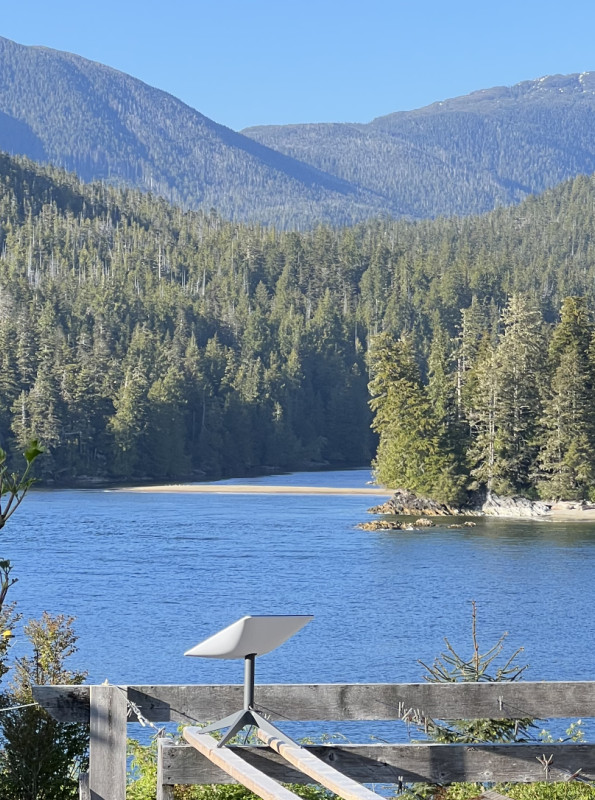}
        \label{fig:koeye_point_dish}
    }
    \subfigure[The surrounding terrain map in Manning Park.]{
        \includegraphics[trim={0 0 0 0},clip,width=.39\linewidth]{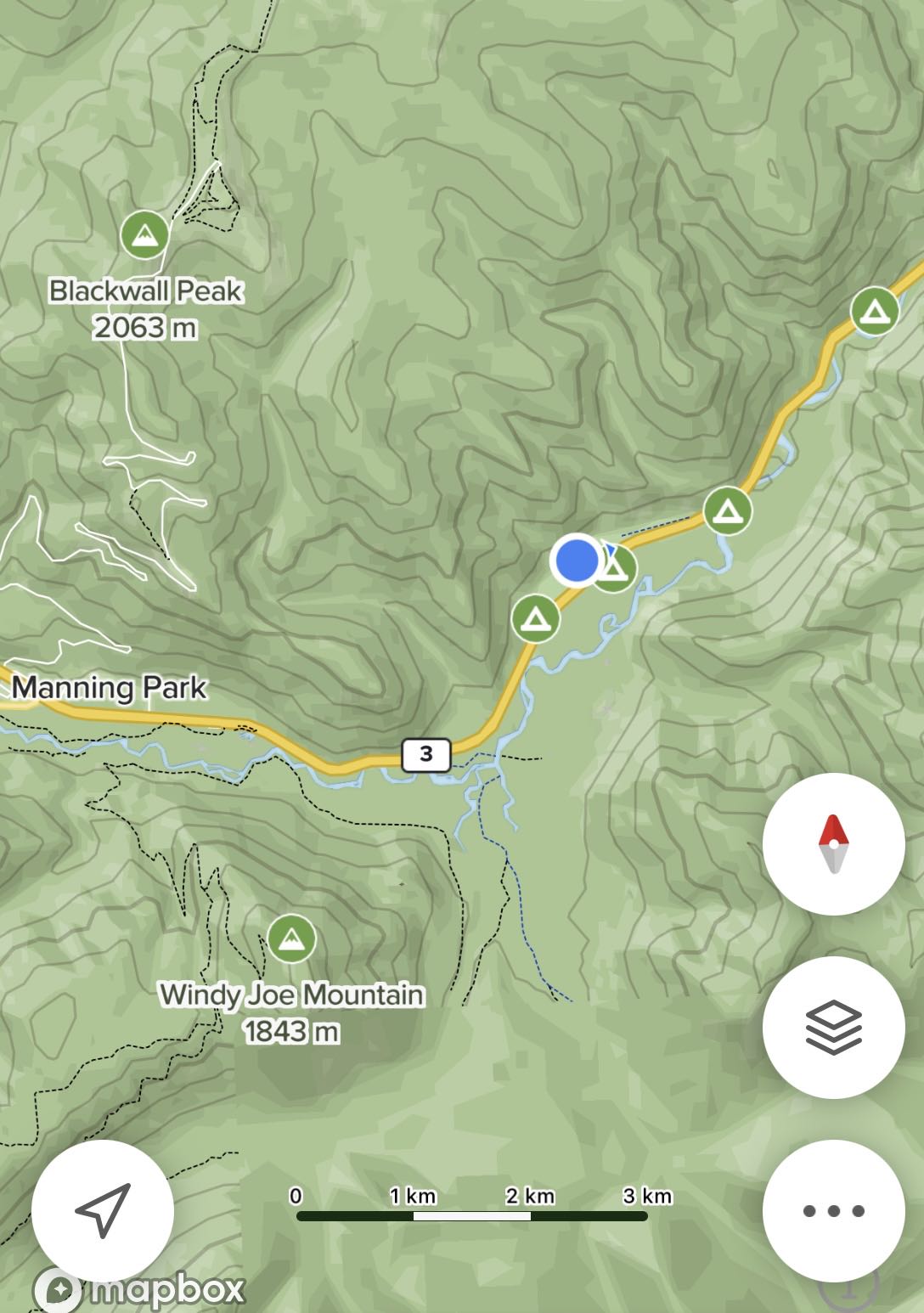}
        \label{fig:manning_park_terrain_map}
    }
    \subfigure[Download throughput of the gen 2 dish at the (R)emote estuary of the Koeye River compared to the gen 1 dish in the (U)rban setting. Vertical lines in the middle of the bars describe the standard deviation.]{
        \includegraphics[trim={0 0 0 0},clip,width=.95\linewidth]{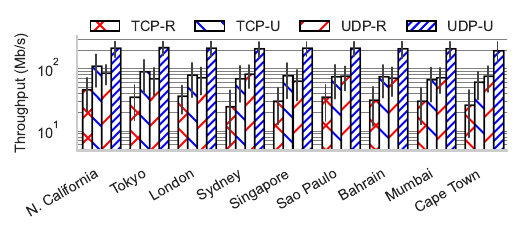}
        \label{fig:koeye_point_down_throughput}
    }
    \caption{The Measurement Setups in Remote Areas}
    \label{fig:koeye_point}
\end{figure}

We evaluated a Starlink second-generation dish at the Koeye River estuary by the Pacific coast (see Fig.~\ref{fig:koeye_point_dish}), ensuring clear sky exposure. Unlike the circular first-generation dishes used in urban experiments, this rectangular model, despite being within Starlink's service range, experienced a significant reduction in download speed—approximately $68\%$ lower than in urban settings across all tested regions (Fig.~\ref{fig:koeye_point_down_throughput}). Latency also increased by $11\%$ to $30\%$, with more frequent disruptions. These issues are likely as a result of sparse satellites and ground stations in northern areas. We expect service improvements as Starlink's network grows and incorporates more inter-satellite links.

\subsubsection{High-Elevation Deep Valley}

Our second assessment occurred in Manning Park, BC, an area surrounded by mountains exceeding 2 kilometers in height, lacking conventional Internet access, as depicted in Fig.~\ref{fig:manning_park_terrain_map}. Despite being marked as non-serviceable at the time, we successfully connected to Starlink. However, latency ranged widely, averaging 90 to 350 ms, with occasional spikes exceeding 1,000 ms. Throughput metrics were constrained, averaging 13 Mbps for downloads and 4 Mbps for uploads, with rare surges to 100 Mbps and 20 Mbps. Despite a consistently low obstruction ratio of $2\%$, service disruptions occurred at intervals of one to three minutes, highlighting the substantial impact of geographic terrain on user-perceived network performance.

\begin{figure}[t]
    \centering
    \subfigure[Koeye River aerial view featuring the 30m tall trees]{
        \includegraphics[width=.7\linewidth]{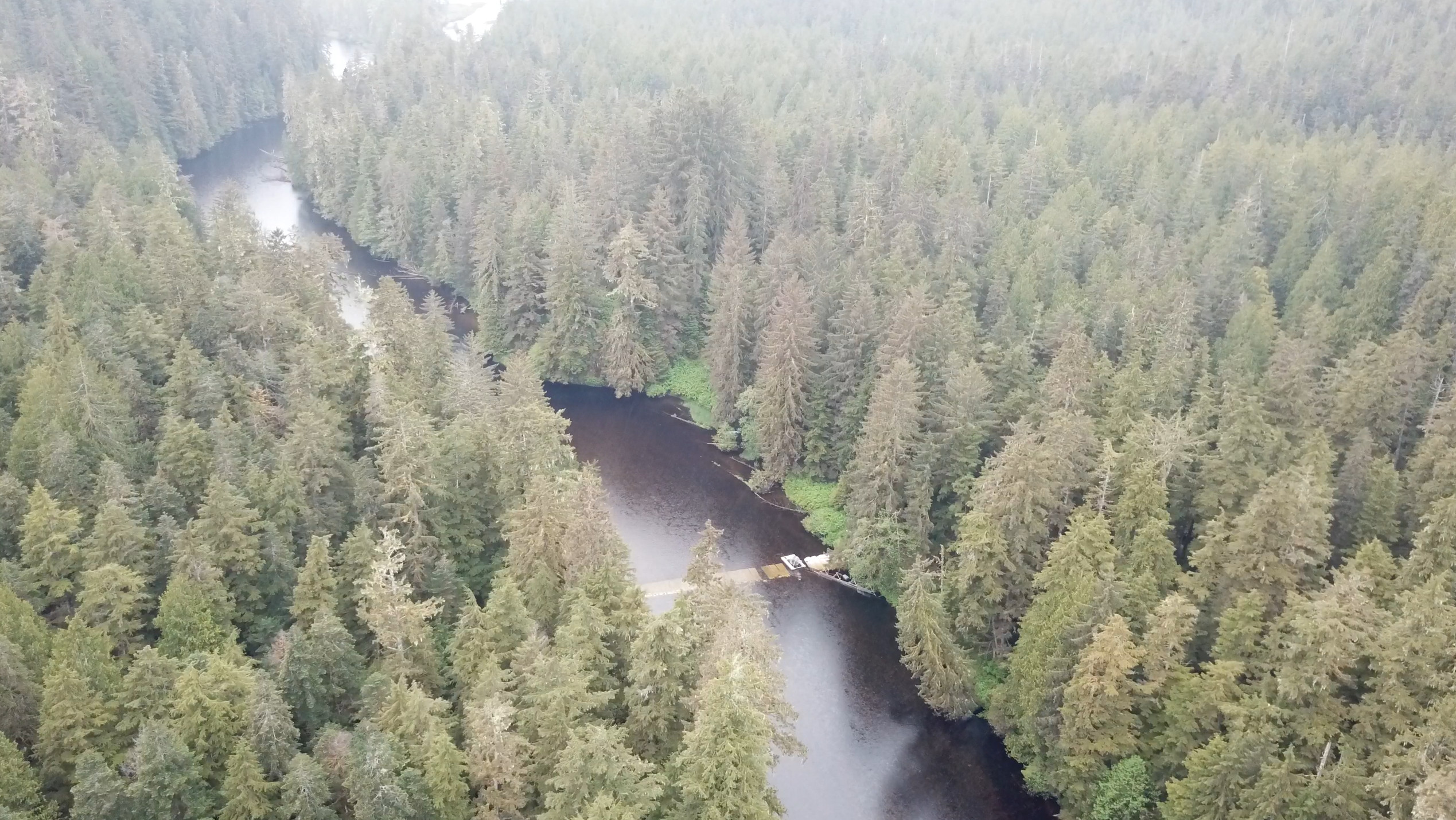}
        \label{fig:koeye_weir_trees}
    }
    \subfigure[Koeye River weir setup and solar panels]{
        \includegraphics[trim={3cm 3cm 3cm 3cm},clip,width=.7\linewidth]{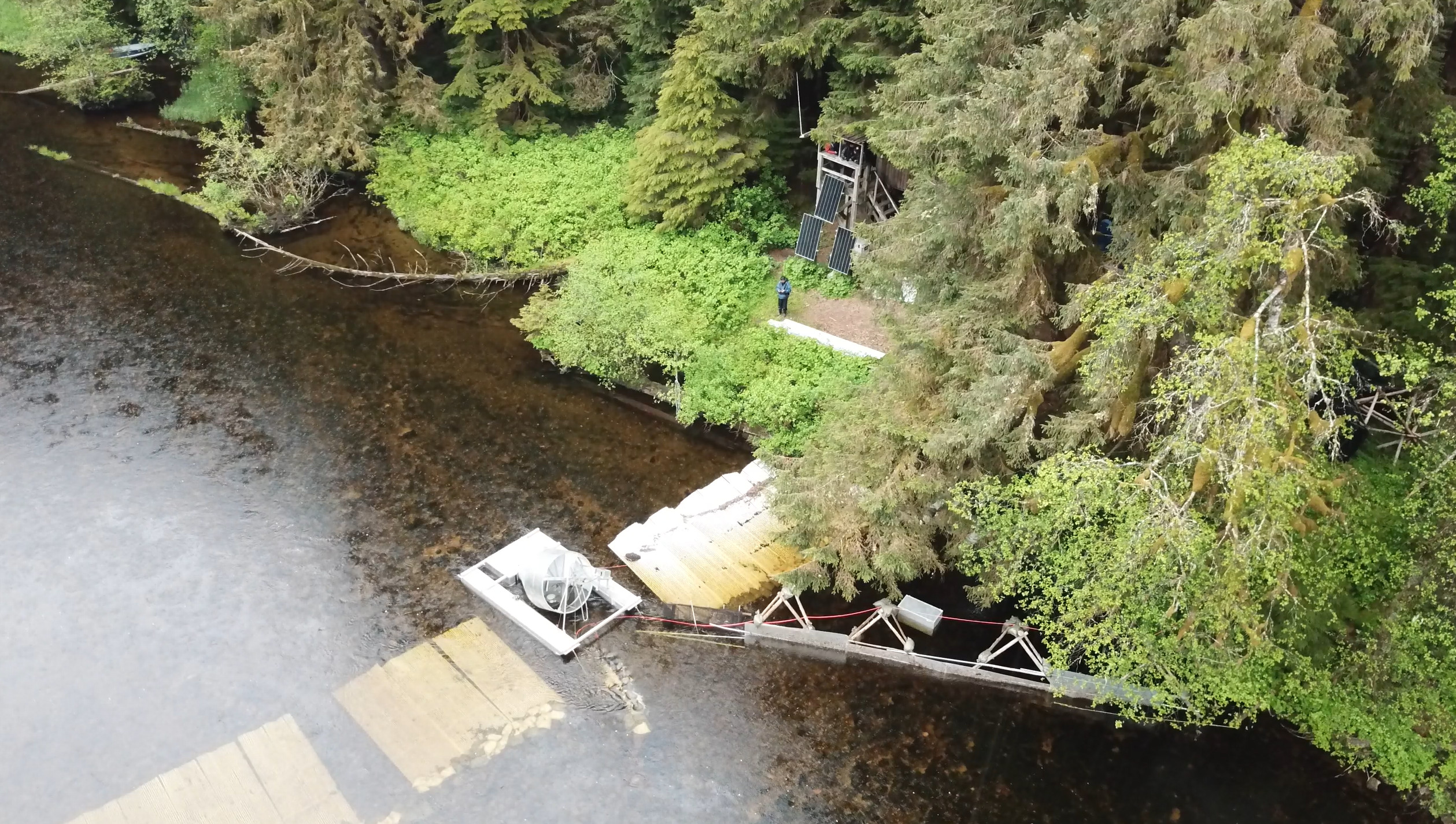}
        \label{fig:koeye_weir_setup}
    }
    \subfigure[Solar panels in the harsh winters of Bear Creek River]{
        \includegraphics[trim={0 0 0 0},clip,width=.7\linewidth]{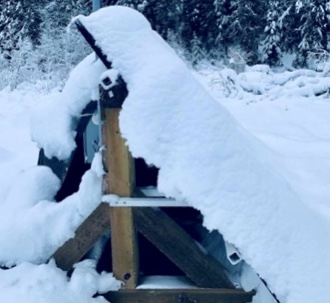}
        \label{fig:winter_solar_panel_bear_creek}
    }
    \caption{An aerial view of the weir site upstream of the Koeye River in the Great Bear rainforest.}
    \label{fig:koeye_weir}
\end{figure}

Current routing strategies in satellite networks, specifically Starlink, primarily employ single bent-pipe transmissions for initial or final segments of data transfer \cite{giuliari_icarus_2021,hauri2020internet}. Our findings indicate that Starlink's performance variability largely stems from the vulnerability of these transmission paths to environmental factors, including solar storms and rainfall. Notably, a solar storm that caused a geomagnetic storm on February 3rd, 2022, resulted in a drastic reduction in Starlink's throughput from 100 Mbps to 5 Mbps, persisting for approximately two days, potentially exacerbated by the atmospheric re-entry of 40 satellites during this period.\footnote{\url{https://www.spacex.com/updates/\#sl-geostorm}, [Online; accessed Mar 25, 2024]} Moreover, our analysis revealed a negative correlation between precipitation levels and network throughput \cite{ma2023network}.

\subsection{Streaming QoE (Quality of Experience)}

Multimedia applications, exemplified by our salmon monitoring system, constitute the majority of Internet traffic. Starlink can facilitate communication for biologists, ending years of isolation during site visits. From an end user's perspective, we initially assess Starlink's support for contemporary commercial multimedia services like YouTube, Twitch, and Zoom conferencing. Our findings indicate satisfactory performance for Video-on-Demand (VoD) applications, benefiting from a large buffer size. In contrast, live streaming services, with smaller buffers, struggle against network dynamics, leading to potential frame drops and frozen playback. In interactive video conferencing, Starlink's higher network latency and jitter significantly impact the user experience, with a $30\%$ increase in average interaction latency compared to terrestrial networks.

\begin{figure}[!t]
    \centering
     \includegraphics[width=\linewidth]{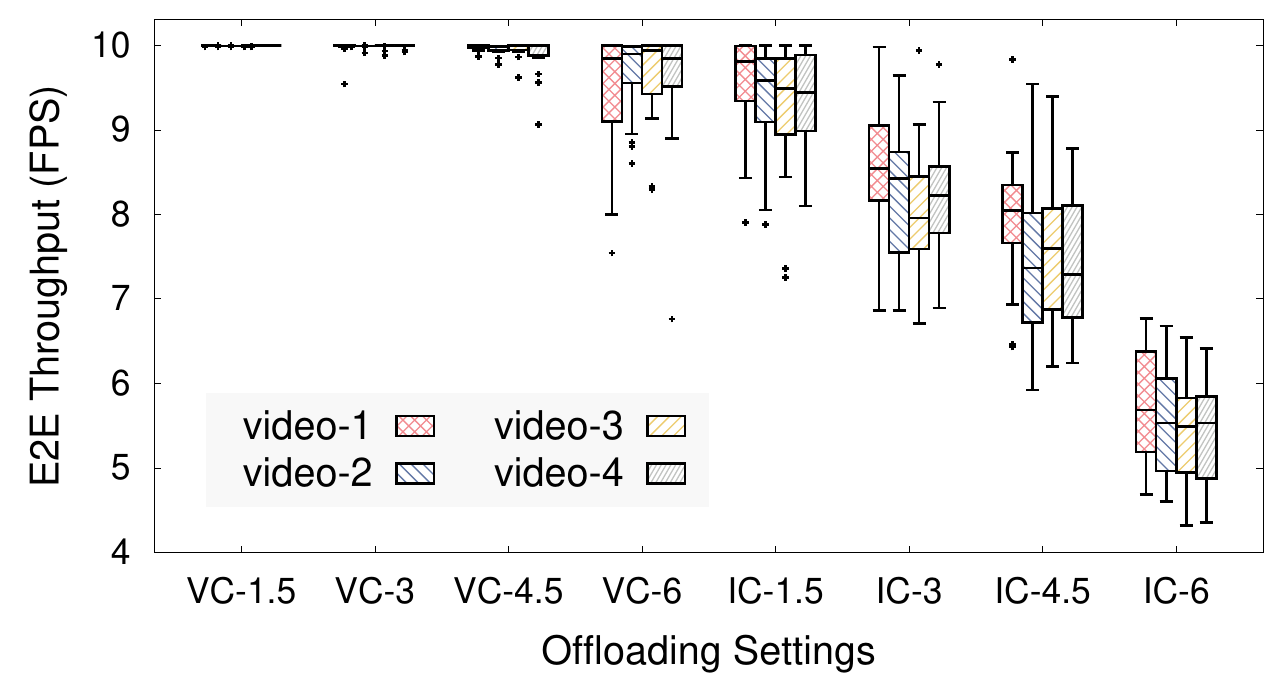}
    \caption{E2E throughput distribution of two video uploading modes over LSN. (The number in X-axis tick labels represents the corresponding bitrate, e.g., VC-6 represents uploading the video encoded at $6$ Mbps.)}
    \label{fig:e2e-throughput}
\end{figure}

Automated salmon monitoring relies on video analytics services, transmitting videos from cameras to remote servers for analysis with advanced learning models. This upload-centric demand conflicts with Starlink's download-centric design. We stream videos at $10$ FPS from a Starlink client to a nearby data center server, considering two uploading modes: 1) uploading h.264 compressed videos with RTMP (VC), and 2) independently uploading JPEG compressed frames one by one over TCP (IC). Fig.~\ref{fig:e2e-throughput} displays the end-to-end throughput distribution over $20$ in-the-wild trials. Supporting RTMP-based video uploading at a higher bitrate, e.g., $6$ Mbps, can be challenging for current Starlink. The significant throughput variation between uploading trials for the same video underscores the substantial impact of underlying network variations on application-perceived QoE. Additionally, supporting medium- to high-frame-rate ($ > 10$ FPS) synchronous offloading for analytics applications with stringent single-frame response requirements, such as AR applications, remains challenging for today's Starlink.

\subsection{Service Dynamics: Impact from Weather and Mobility}

Our investigation shows that Starlink's throughput is weather-sensitive, exhibiting an inverse correlation with precipitation and temperature. On average, throughput declines by $27\%$ during any form of precipitation \cite{ma2023network}. UDP downloads, averaging 215 Mbps in dry conditions, drop to approximately 120 Mbps during heavy rainfall ($>4$ mm per hour).\footnote{\url{https://water.usgs.gov/edu/activity-howmuchrain-metric.html}, [Online; accessed Sep 27, 2023]} This reduced throughput is likely due to rain attenuation, with Ka- and Ku-band radio waves being sensitive to rainfall \cite{panagopoulos_satellite_2004, kourogiorgas_space-time_2018}. Additionally, even lighter cloud formations can attenuate satellite signal strength by approximately $10\%$ \cite{vasisht_l2d2_2021}.

Thunderstorms also significantly impact Starlink's performance, causing 420\% more frequent and prolonged network outages compared to clear days. These extended outages render the Starlink network unstable during adverse weather, posing challenges for our deployment in temperate rainforests, known for their humidity and cloudiness. This condition contributes to consistently lower bandwidth compared to measurements in urban areas.

Starlink's recent FCC authorization for in-vehicle usage opens possibilities for portable and dynamic applications.\footnote{\url{https://www.cnbc.com/2022/06/30/fcc-approves-spacex-starlink-service-to-vehicles-boats-planes.html}, [Online; accessed Sep 27, 2023]} While mobility features may not directly impact salmon monitoring, in-transit connectivity can benefit maintenance personnel moving from urban to remote areas for system assessment. This can aid in deploying complex alert systems or maintaining seamless communication with emergency personnel. To evaluate Starlink's performance for mobile scenarios, we conducted a 30-minute test near the Coquitlam River, a significant salmon hatchery site. A Starlink dish, securely fastened to a minivan roof rack, was tested with the portability option activated.\footnote{This test was conducted for academic research purposes only. Starlink currently does not endorse dish use on moving vehicles, but trials are underway, notably in Ukraine, and an RV version has been released.}

While functional in a mobile setting, Starlink requires substantial improvements to match the reliability of stationary setups. The system encountered frequent outages—every 16.5 seconds—lasting 5 to 36 seconds, with an average latency of 100-200 ms. These outages may have resulted from satellite handovers, confirmed every 15 seconds \cite{tanveer2023making}, aligning with our observations, and the dish's mobility could have contributed to longer search times for new satellites during these handovers. Download throughput ranged from 80-100 Mbps but rapidly diminished, while upload speeds remained consistent at 5-7 Mbps, akin to stationary setups. These latency and throughput shortcomings significantly impact service quality, requiring further refinement.

Additionally, the Starlink kit generally takes between 3-7 minutes to initialize and establish an Internet connection, but we observed instances where this duration exceeded 20 minutes. This extended boot-up time could serve as a significant barrier to effective mobile utilization.

\subsection{The Wild World Challenge}

While technical issues such as latency and throughput are likely to improve over time, our expedition to the upstream areas of the Great Bear temperate rainforest highlighted a range of pragmatic challenges that warrant close attention. Our team was the first external group to visit the weir at the Koeye River in two years, illustrating the remote nature of the site.

\subsubsection{Environmental Constraints}
The towering old-growth trees surrounding the river, as depicted in Fig.~\ref{fig:koeye_weir_trees}, rise to heights of at least 30 meters and restrict clear sky visibility to the riverside weir. This area is also prone to flooding. Consequently, the Starlink dish at KwaKwa was mounted 25-30 feet above ground level to avoid tree obstructions, while the Koeye dish was situated in the middle of the river on a fish weir, reconstructed based on archaeological evidence from the Heiltsuk Indigenous community (see Fig.~\ref{fig:koeye_weir_setup}). The uptime for the KwaKwa setup significantly underperformed compared to the Koeye dish, incurring outages every 5-10 minutes for 2-3 minutes, likely due to the obstructive foliage.

\subsubsection{Wildlife Considerations}
The local ecosystem, rich in bears, wolves, and various avian species, introduces risks to the stability of the dish setup. Specifically, the Koeye watershed is home to an estimated population of 100 individual grizzly bears. Wildlife can easily travel through the narrow weir, posing a potential threat to the equipment.

\subsubsection{Power Considerations}
Given the remote location, power is primarily sourced from solar panels, which operate under the limitations of daylight and weather conditions. The solar arrays produce up to 990W per hour under optimal conditions while the Starlink dish's power consumption averages at 56.3 Watts and peaks to 144.5 Watts\cite{ma2023network}. Our estimations of the entire system suggest that 3-4 hours of direct sunlight per day would suffice for continuous operation; however, fog and rainfall led to periodic outages despite battery backup systems capable of providing up to 48 hours of power. Winter conditions further impede solar power generation, as shown in Fig.~\ref{fig:winter_solar_panel_bear_creek}, where snow cover must be manually cleared to resume energy capture for the nearly sustained higher power requirements of 145 Watts the dish uses to maintain snow melting mode\cite{ma2023network}.

\subsubsection{Logistical Constraints}
The isolated nature of the location makes diesel generators impractical and unauthorized, leaving us dependent on utilizing solar panels. Moreover, access to the area upstream of the Koeye River is tidal-dependent, only during high tides, making physical maintenance costly and sometimes impossible.

\section{Towards Ubiquitous LSN Access}

\subsection{Cross-Orbit Collaboration}

\begin{figure}[t]
    \centering
     \includegraphics[width=\linewidth]{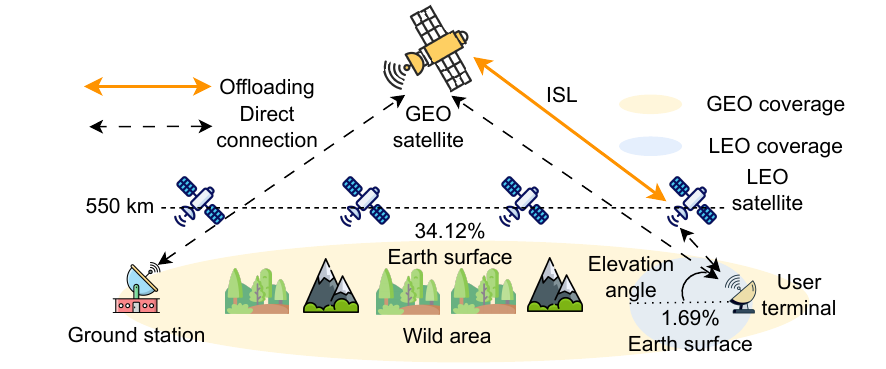}
    \caption{Cross-orbit collaboration in wild area.}
    \label{fig:cross_orbit_collaboration}
\end{figure}

Due to the fast growth of the LSN market, the number of LEO satellites and operators surges. This fast growth trend will lead to a crowded LEO space with a massive number of LEO satellites. We envision that satellite networks will be similar to the Internet today, where operators, including LEO, {\em Medium Earth Orbit} (MEO), and GEO, collaborate and share cross-orbit resources, providing end users with cross-orbit services in a single terminal without the need for additional satellite deployments. Intelsat, for instance, is testing their cross-orbit services using electronically steered arrays to switch services between LSNs and GEO satellite networks.\footnote{\url{https://www.intelsat.com/newsroom/intelsat-completes-multi-orbit-inflight-wi-fi-tests/}, [Online; accessed Sep 26, 2023]}

Cross-orbit collaboration is promising to improve service continuity in remote areas to address LEO coverage limitations. As shown in Fig. \ref{fig:cross_orbit_collaboration}, a single LEO satellite at a $550$ km altitude and $10$\textdegree{} minimum elevation angle covers just $1.69\%$ of Earth's surface, insufficient for comprehensive remote area and ground station connectivity. In contrast, a GEO satellite can cover $34.12\%$ of the surface, facilitating direct end-user connections and enabling LEO-GEO traffic offloading for broader reach. This approach is particularly beneficial for polar region operations, where existing networks like Starlink, with a $53$\textdegree{} inclination, predominantly serve lower latitudes, neglecting polar coverage. Cross-orbit collaboration can extend LSNs reach to polar areas through GEO satellite network integration.

The costs of the operators who sell their resources must be compensated. However, since the buyers only have a limited budget, a pricing algorithm must be carefully designed and integrated with the routing algorithm. Our preliminary evaluation has shown that our pricing and routing algorithm can conserve $11.00\%$ to $63.85\%$ more energy on satellites and reduce $33.90\%$ to $81.56\%$ more delay compared to the routing schemes that did not consider pricing and only focused on transmission rate, delay, and current battery capacity.\footnote{The evaluation results are derived from the authors' ongoing research, which is currently in the process of being submitted to future conferences.}

To mitigate the service quality drops in single GEO offloading scenarios, cross-orbit collaboration enables multiple satellites to collaborate with LEO satellites using a reverse-auction framework, where collaborating operators propose bids detailing the traffic amount they can handle. The LEO satellite operator then offloads traffic to selected satellites, ensuring that no collaborating satellite is overwhelmed.

\subsection{Coordinated Multi-Path Transmission with ISL}

\begin{figure}[t]
    \centering
     \includegraphics[width=\linewidth]{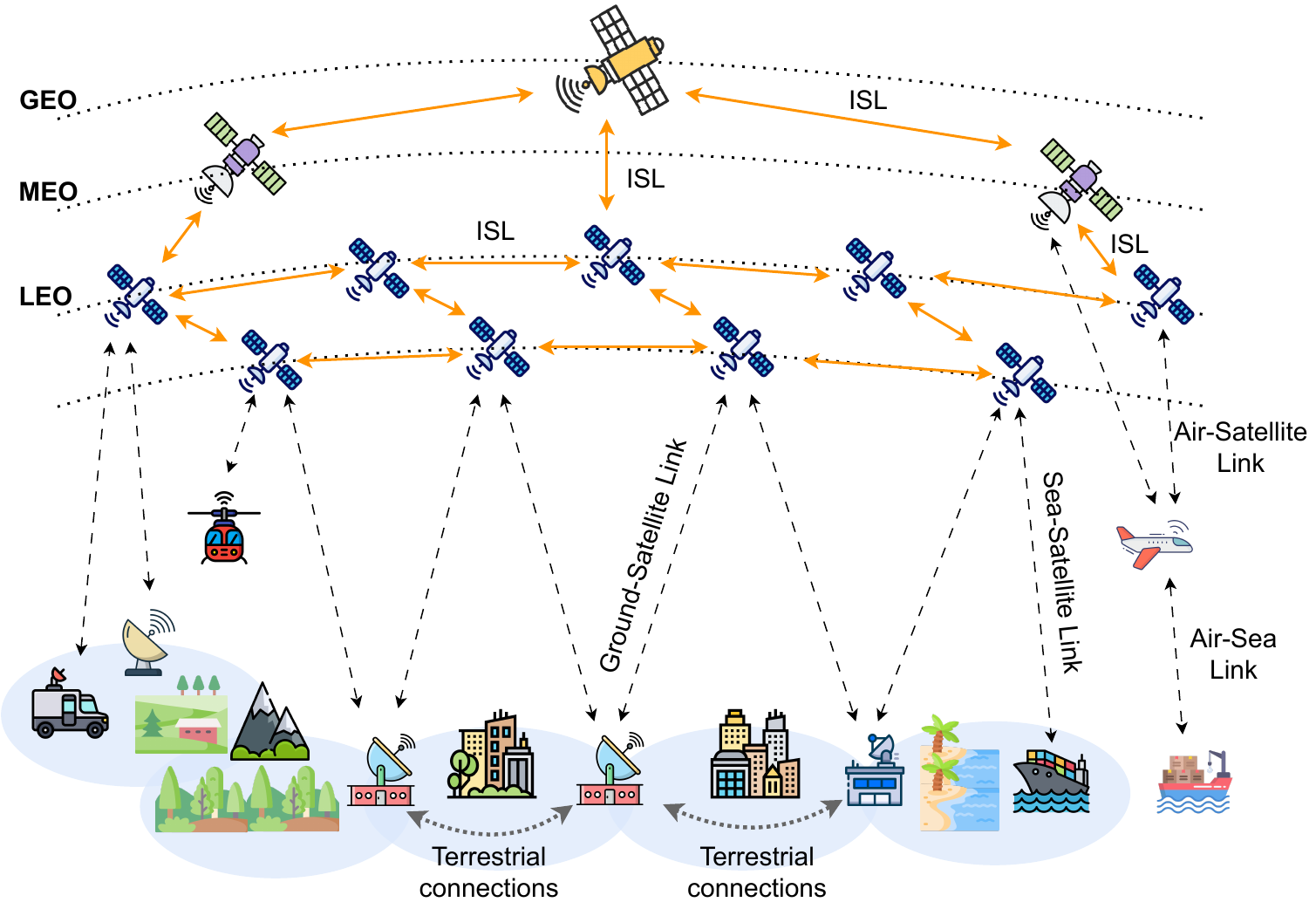}
    \caption{Ubiquitous access powered by coordinated communications.}
    \label{fig:isl}
\end{figure}


In the early stages of Starlink deployment, each LEO satellite served as a traffic relay, directly forwarding received end user traffic down to a GS \cite{lai2023starrynet}. Such bent-pipe style communications require massive GSes to achieve truly global coverage, which contradicts Starlink's goal of reducing the reliance on ground infrastructures. Although there are already about $150$ operational Starlink GSes,\footnote{\url{https://starlinkinsider.com/starlink-gateway-locations/}, [Online; accessed Sep 24, 2023]} it is still far from being satisfactory to achieve ubiquitous LSN access for end users. The newer generation of Starlink satellites, version 1.5 and 2.0, are equipped with optical heads for inter-satellite communications.\footnote{\url{https://en.wikipedia.org/wiki/Starlink}, [Online; accessed Sep 24, 2023]} ISLs provide alternative network paths in space with higher reliability and data fidelity, significantly reducing the signal attenuation and latency variations caused by weather conditions \cite{hauri2020internet}. As more satellites become interconnected via ISLs, a new form of low-latency, wide-area backbone that connects any end user anywhere is emerging.

As shown in Fig.~\ref{fig:isl}, the complexity of the network topology multiplies with integrated ISLs, ground-satellite links, and different types of ground and air relays. A number of paths with varying delays, costs, and throughputs can be available for any pair of end users. Coordinating the multi-path transmission can increase network reliability and security, improve in-the-wild performance, and facilitate load balancing. For example, multiple ISL paths could be invoked simultaneously for transoceanic multimedia transmission to satisfy the low latency and high throughput requirements. For remote areas that experience frequent power outages, alternative ISL paths can provide resilience to terrestrial paths and help maintain connectivity. Yet, challenges need to be addressed to realize efficient coordination, such as data packet synchronization and ordering, topology- and performance-aware dynamic path selection, end-to-end multipath device compatibility, and standardization for interoperability.

\subsection{LSN Empowered Mobile Edge Computing}

As LSN evolves into seamless global coverage, every end user will have equal access to the Internet and enjoy the convenience of modern technologies, such as online gaming and video streaming. Emerging applications that are hard to develop with terrestrial networks will be easily implemented, such as wildlife monitoring, disaster response and relief, and maritime surveillance. Typically, these applications require edge computing resources to deliver satisfactory QoE such as low latency. As each satellite in the LSN can directly communicate with end user devices at a latency as low as $25$ ms,\footnote{\url{https://www.starlink.com/technology}, [Online; accessed Sep 23, 2023]} LEO satellites are the ideal mobile edge computing (MEC) nodes if they are equipped with sufficient onboard computing resources. Given the large coverage size of GEO and MEO satellites, they will likely receive excessive service requests beyond their computing capacity when satellite access becomes ubiquitous. With cross-orbit communication, the dense LEO MEC nodes may help by sharing fractional workloads offloaded from GEO or MEO satellites while improving their own onboard computing resource utilization, especially when flying over areas with rare end user requests.

Unlike terrestrial MEC, where the edge computing nodes (e.g., base stations) are stationary, LEO MEC needs to address the challenges originating from both the end users and edge computing nodes. This is because the LEO satellites move relatively faster than the Earth, completing one full orbit around the Earth typically in $90$-$120$ minutes.\footnote{\url{https://aerospace.csis.org/aerospace101/earth-orbit-101/} [Online; accessed Sep 23, 2023]} The high mobility results in frequent satellite handovers, occurring around every $15$ seconds, according to a recent study \cite{tanveer2023making}. The frequent handovers can cause a series of issues, including unexpected service interruptions, fluctuations in available computing resources, potential data privacy leakage, and latency variations \cite{cao2023satcp}. This calls for mobility-aware cross-satellite resource scheduling and task dispatching strategies that consider the satellite movement trajectories, available onboard computing resources, and QoE requirements of edge computing tasks.

Compared with terrestrial MEC nodes, LEO satellites have limited onboard computing resources due to the size, weight, and power constraints. Although LEO MEC nodes can take on partial latency-sensitive computing tasks, most heavy-lifting workloads still need to rely on terrestrial computing infrastructures. Fortunately, recent years have witnessed the accelerated integration of LSN and existing terrestrial infrastructures. For instance, Starlink has partnered with Google Cloud to install GSes within or near cloud data centers to directly ingest user data from space at the edge of LSN \cite{kassem2022browser}. Other cloud providers also successively roll out Ground Station as a Service (GSaaS) offerings, such as AWS Ground Station,\footnote{\url{https://aws.amazon.com/ground-station/}, [Online; accessed Sep 23, 2023]} to provide available ground antenna and geo-distributed computing infrastructures for user data from space. The integrated terrestrial-satellite computing infrastructures make solid foundations for ubiquitous and responsive LSN access.

\section{Further Opportunities and Challenges of Global Coverage with LSN}

Acknowledging the essential role of the LSN in attaining global coverage, the Starlink dish acted as the crucial bridge between remote areas and the broader world. Therefore, it is imperative to assess the current alternative networks in these remote regions and consider the potential challenges associated with deploying the Starlink dish. It is also necessary to examine the industrial and governmental policies, as well as the social and cultural implications.

\subsection{Opportunities for LSN: Africa as a case}

In comparison to other regions, Africa's network infrastructure is comparatively underdeveloped. In the US, Canada, and most East Asian nations, more than 80\% of the population has access to the Internet. However, only large cities in Africa have access to 4G networks, and only 39.7\% of the population has internet access.\footnote{\url{https://en.wikipedia.org/wiki/List_of_countries_by_number_of_Internet_users}, [Online; accessed Sep 26, 2023]} 

To evaluate the current performance of terrestrial networks and uncover potential advantages of LSN in these remote areas, we have conducted a preliminary measurement on cellular network performance in Gabon, Africa as a case study. One test site was situated in a village approximately 25 kilometers away from the nearest city, Franceville. We used an online speed test website\footnote{\url{https://pcmag.speedtestcustom.com/}, [Online; accessed Sep 26, 2023]} and selected a server located in Franceville for our testing purposes. The average download and upload bandwidth were measured at 4.10 and 0.27 Mb/s respectively, with a latency of 303 ms and jitter of 89 ms. It is worth noting that for the experiment, a signal amplifier and an LTE router are necessary for Internet access. Otherwise, standard mobile phones even struggled to receive the signal. In comparison to the current cellular network available in remote parts of Africa, the LSN services that we have experienced are far superior, albeit yet to cover this vast continent.

There are challenges associated with deploying Starlink dishes in this area. The Starlink dish is known to be sensitive to extreme weather conditions, including high temperatures and precipitation\cite{ma2023network}. Given that Africa has a tropical wet and dry climate characterized by high temperatures and frequent heavy precipitation, it is anticipated that the Starlink dish may experience performance degradation. As an example, Franceville experiences annual temperature variations between 20-31\degree C,\footnote{{\url{https://weatherspark.com/y/148278/Average-Weather-at-Franceville-Mvengue-Airport-Gabon-Year-Round}, [Online; accessed Sep 26, 2023]}} potentially leading to a throughput decline of 5-26\% due to the absence of a positive cooling mechanism for the Starlink dish\cite{ma2023network}. Additionally, the year 2022 saw approximately 43\% of Africa's population without access to electricity.\footnote{\url{https://www.iea.org/reports/africa-energy-outlook-2022/key-findings}, [Online; accessed Sep 26, 2023]} Failing to secure a sufficient electrical power supply could present an additional potential challenge. For instance, our test site is outfitted with multiple electric generators that operate exclusively during daytime hours to minimize costs. Despite this, the power capacity remains inadequate to support the simultaneous operation of all high-power electrical appliances.

\subsection{Political Considerations from Industry and Government}

On top of the already approved 12,000 satellites, the FCC has greenlit Starlink's plan to deploy an additional 7,500 second-generation satellites.\footnote{\url{https://www.cnbc.com/2022/12/01/fcc-authorizes-spacex-gen2-starlink-up-to-7500-satellites.html}}\textsuperscript{,}\footnote{\url{https://www.space.com/spacex-starlink-satellites.html}} Starlink employs higher-frequency Ka- and Ku-bands for gateway and user communications respectively \cite{leo_network_updated_2021}, optimizing for higher data rates but has increased vulnerability to atmospheric and rain attenuation. This necessitates the use of high-power transmissions and high-gain antennas with narrow beam widths.

Given that orbital and spectral resources are limited, the International Telecommunication Union has implemented guidelines to ensure equitable access and resource efficiency. The allocation for LEO orbit and spectrum resources operates on a ``first-come, first-served'' basis with a time limit of 7 years being imposed for notifying and putting the system into use.\footnote{\url{https://www.itu.int/en/ITU-R/space/snl/Documents/ITU-Space_reg.pdf}, [Online; accessed Sep 25, 2023]} Failure to use the reserved spectrum within this time frame renders it invalid. These regulations provide substantial advantages to pioneers, represented by Starlink, enabling them to maintain a dominant position within this segment. The latecomers, however, have to address the scarce orbit and spectrum resources challenge and the operation and coordination challenge of not creating interference with existing services.

To achieve truly global coverage, Starlink incorporates laser-based Inter-Satellite Links (ISLs) to bypass limitations such as small individual satellite coverage areas and a lack of GSes in certain regions\cite{handley2018delay}. ISLs facilitate data transmission at the speed of light in a vacuum, approximately 47\% faster than through optical fiber, thus enabling lower-latency, long-distance communications. The adoption of ISLs however has elicited significant regulatory concerns, primarily centered around data security. Authorities are apprehensive that the ISL-based routing could lead to a lack of control over data traffic, raising questions about the integrity and security of data as it transits through space.\footnote{\url{https://spacenews.com/oneweb-says-regulatory-concerns-main-reason-its-forgoing-inter-satellite-links/}, [Online; accessed Sep 22, 2023]}

\subsection{Environmental, Social and Cultural Implications}

While the technological advancements of LSNs are widely celebrated, it is imperative to address their broader environmental, social, and cultural implications, many of which are not yet fully understood.

One significant concern is the escalation of space debris, a mixture of natural meteoroids and man-made orbital fragments, which poses substantial risks to operational satellites and spacecraft. Collisions with even minuscule fragments can lead to catastrophic outcomes, as past incidents have illustrated \cite{european2010iridium}. Furthermore, the increasing saturation of LEO with both operational satellites and debris poses challenges for astronomers. These objects, when illuminated by sunlight, can hinder telescopic observations by falling within the 4th to 6th visible magnitude range, thereby impacting the quality of astronomical research \cite{light_pollution}. To address these challenges, Starlink's second-generation (V2.0) satellites have undergone design modifications aimed at reducing their luminosity and mitigating their impact on astronomical activities.\footnote{\url{https://api.starlink.com/public-files/BrightnessMitigationBestPracticesSatelliteOperators.pdf}, [Online; accessed Sep 25, 2023]}

The sociocultural challenges are no less complex, a fact brought to light during our field studies. For instance, the Koeye river and its surrounding rainforest hold spiritual significance for the Heiltsuk Indigenous community, and the introduction of modern values into such sacred spaces has incited conflicts,\footnote{\url{https://ojs.library.ubc.ca/index.php/bcstudies/article/view/1913}, [Online; accessed Oct 04, 2023]} symbolizing a clash between traditional values and technological advancements. This will potentially cause tension between traditional cultural values and the upcoming modern technological connectivity.

Moreover, the idea of signal pollution extends beyond technological aspects into ethical and philosophical domains. The advent of ubiquitous connectivity is not without its drawbacks, as it also contributes to a new form of intangible pollution: {\em WiFi pollution}. This intrusion into once-pristine areas has raised concerns among biologists, who argue that the distractions of mobile applications could compromise the attentiveness and self-reliance essential for survival in wilderness settings.

In conclusion, the quest for global internet coverage necessitates the formulation of interdisciplinary policies, professional guidelines, and ethical standards. These comprehensive frameworks must be sensitive to the array of impacts—technological, environmental, social, and cultural—posed by the proliferation of LSNs.

\section{Conclusions}
\label{sec:conclusion}

This article has presented our initial experience with Starlink's network access in remote areas. Our experience has suggested that a space Internet based on LEO satellite constellations has great potentials towards global coverage, particularly as a compliment to cellular services. There are however challenges from industry and government regulations, and from environmental, social and cultural considerations. We discussed a series of potential solutions and, in spite of the challenges, we expect that truly ubiquitous network access for general users would soon be realized through a seamlessly integration of diverse terrestrial and space networks. 

\section*{Acknowledgment}
This project was supported by a Canada NSERC Discovery Grant and a British Columbia Salmon Recovery and Innovation Fund (No. 2019-045). The authors thank Ian Clevenger from the Salmon Watersheds Lab at SFU, as well as the Heiltsuk First Nation, for their great support. We also thank the QQS (EYES) Projects Society for hosting us at the Koeye Lodge (in particular, the lodgekeepers Ian and Emily Files) and for their effort in protecting the natural environment of the Great Bear Rainforest and the Heiltsuk heritage.

\bibliographystyle{IEEEtran}
\bibliography{leo_wild_magazine}

\section*{Biographies}

\begin{IEEEbiographynophoto} {Sami Ma} received his B.Sc. degree with distinction in Computing Science at Simon Fraser University, BC, Canada in 2019. Currently, he is continuing doctoral studies in Computing Science at Simon Fraser University. His research interests include low earth orbit satellite networks, internet architecture and protocols, deep learning, and computer vision.
    
\end{IEEEbiographynophoto}

\begin{IEEEbiographynophoto} {Yi Ching Chou}
received his B.B.A. in finance and B.Sc. in computing science degrees from Simon Fraser University, British Columbia, Canada. He is currently a Ph.D. student in computing science at Simon Fraser University and received the NSERC CGS M award. His research interests include satellite communications, computer networking, and edge computing.
    
\end{IEEEbiographynophoto}

\begin{IEEEbiographynophoto}{Miao Zhang}
received her B.Eng. degree from Sichuan University, Chengdu, China, in $2015$, and her M.Eng. degree from Tsinghua University, Beijing, China, in $2018$. She is currently a Ph.D. student at Simon Fraser University, British Columbia, Canada. Her research areas include cloud and edge computing, and multimedia systems and applications.
\end{IEEEbiographynophoto}

\begin{IEEEbiographynophoto} {Hao Fang}
    received his B.Sc. (Hons.) degree with distinction in Computing Science from Simon Fraser University, BC, Canada, in $2022$. He is currently a Ph.D. student in Computing Science at Simon Fraser University. His research areas include satellite communications and networking, particularly with multimedia systems.
\end{IEEEbiographynophoto}

\begin{IEEEbiographynophoto} {Haoyuan Zhao} 
received his B.Sc. in computing science degrees from Simon Fraser University, British Columbia, Canada. He is currently a Ph.D. student in computing science at Simon Fraser University. His research interests include satellite networking and multimedia systems. 

\end{IEEEbiographynophoto}

\begin{IEEEbiographynophoto}{Jiangchuan Liu} (S'01-M'03-SM'08-F'17) is a Professor at Simon Fraser University, BC, Canada. He is a Fellow of The Canadian Academy of Engineering and an IEEE Fellow. He received BEng (cum laude) from Tsinghua and PhD from HKUST. He has served on the editorial boards of IEEE/ACM Transactions on Networking, IEEE Transactions on Multimedia, IEEE Communications Surveys and Tutorials, and IEEE Internet of Things Journal. He was a Steering Committee member of IEEE Transactions on Mobile Computing and Steering Committee Chair of IEEE/ACM IWQoS. He was TPC Co-Chair of IEEE INFOCOM'2021 and General Co-Chair of INFOCOM'2024.
\end{IEEEbiographynophoto}

\begin{IEEEbiographynophoto} {William I. Atlas} joined the Wild Salmon Center in 2020. Prior to joining, he spent ten years in British Columbia, working with First Nations to build community-based salmon science initiatives. This research has focused on population monitoring and assessment tools for remote watersheds on the Central Coast of BC, and understanding how fisheries and climate act synergistically to drive salmon population trajectories. He holds an MSc and PhD in Biological Sciences from Simon Fraser University, and a BSc in Aquatic and Fishery Sciences from the University of Washington.
    
\end{IEEEbiographynophoto}

\end{document}